%% file: manuscript_uRWELL_PICOSEC.tex
\documentclass[3p,times]{elsarticle}
\usepackage[breaklinks,hidelinks,colorlinks]{hyperref}

\usepackage{mathtools}
\usepackage{enumitem}
\usepackage{natbib}
\usepackage{soul}
\usepackage[dvipsnames]{xcolor}
\usepackage[T1]{fontenc}
\makeatletter
\renewcommand{\fnum@figure}{Fig. \thefigure}
\makeatother
\usepackage{lineno}
\begin{document}
\begin{frontmatter}
%
%
%
\title{\texorpdfstring{$\muup$}{}RWELL-PICOSEC: Precision Timing with Resistive Micro-Well Detector}
%
%
%
%
\input{00_authorList}
\begin{abstract}

The PICOSEC detector concept uses a micro-pattern gaseous detector (MPGD) amplification structure combined with a Cherenkov radiator coated with a semi-transparent photocathode to provide below tens of picosecond-level precision timing capabilities with minimum ionizing particles. PICOSEC has triggered interest in the development of time-of-flight detectors for particle identification and timing detectors for track reconstruction in the high rate environment of future nuclear and high energy physics experiments. The PICOSEC Micromegas (or PICOSEC-MM) detector, developed by the CERN-based PICOSEC collaboration, use the Micromegas structure for gaseous amplification and achieve below 20 ps timing resolution. A new type of PICOSEC detector, the $\muup$RWELL-PICOSEC based on $\muup$RWELL amplification structure, is being investigated at Thomas Jefferson National Accelerator Facility (Jefferson Lab) alongside PICOSEC-MM R\&D efforts in Europe. Preliminary results from the two 2024 beam test campaigns at CERN demonstrate a timing performance better than 24 ps is achievable with a single-channel $\muup$RWELL-PICOSEC prototype. A vigorous R\&D effort is ongoing to improve the timing performance, robustness and operational stability of $\muup$RWELL-PICOSEC detectors. Development of a large size $\muup$RWELL-PICOSEC is also under consideration for applications in large scale experiments.
\end{abstract}
\end{frontmatter}

%
%
\input{01_introduction}
\input{02_small_proto}
\input{03_testbeam_setup}
\input{04_results}
\input{05_conclusion}
\input{06_acknowledgement}
%
\bibliographystyle{elsarticle-num}
\bibliography{ref.bib}
\end{document}

%% file: 00_authorList.tex
\author[inst10]{K Gnanvo\texorpdfstring{\corref{cor1}}{}}
\cortext[cor1]{Corresponding author:}
\texorpdfstring{\ead{kagnanvo@jlab.org}}{}
\author[inst3]{R. Aleksan}
\author[inst4,inst5]{Y. Angelis}
\author[inst3]{S. Aune}
\author[inst6]{J. Bortfeldt}
\author[inst1]{F. Brunbauer}
\author[inst7,inst8]{M. Brunoldi}
\author[inst4,inst5]{E. Chatzianagnostou}
\author[inst9]{J. Datta}
\author[inst10]{K. Dehmelt}
\author[inst11]{G. Fanourakis}
\author[inst1]{S. Ferry}
\author[inst7,inst8]{D. Fiorina\texorpdfstring{\fnref{fnref{fn1}}{}}}
\author[inst1,inst12]{K. J. Floethner}
\author[inst13]{M. Gallinaro}
\author[inst14]{F. Garcia}
\author[inst3]{I. Giomataris}
\author[inst3]{F.J. Iguaz\texorpdfstring{\fnref{fnref{fn2}}{}}}
\author[inst1]{D. Janssens}
\author[inst3]{A. Kallitsopoulou}
\author[inst5]{I. Karakoulias}
\author[inst16]{M. Kovacic}
\author[inst10]{B. Kross}
\author[inst17]{C.C. Lai}
\author[inst3]{P. Legou}
\author[inst1]{M. Lisowska}
\author[inst18]{J. Liu}
\author[inst12,inst19]{M. Lupberger}
\author[inst1,inst4]{I. Maniatis\texorpdfstring{\fnref{fnref{fn3}}{}}}
\author[inst10]{J. McKisson}
\author[inst18]{Y. Meng}
\author[inst24]{M. Micetic}
\author[inst1,inst19]{H. Muller}
\author[inst1]{R. De Oliveira}
\author[inst1]{E. Oliveri}
\author[inst1,inst20]{G. Orlandini}
\author[inst10]{A. Pandey}
\author[inst3]{T. Papaevangelou}
\author[inst21]{M. Pomorski}
\author[inst1,inst22]{M. Robert}
\author[inst1]{L. Ropelewski}
\author[inst24]{K. Salamon}
\author[inst4,inst5]{D. Sampsonidis}
\author[inst1]{L. Scharenberg}
\author[inst1]{T. Schneider}
\author[inst21]{E. Scorsone}
\author[inst9]{N. Shankman}
\author[inst1,inst3]{L. Sohl\texorpdfstring{\fnref{fnref{fn4}}{}}}
\author[inst1]{M. van Stenis}
\author[inst23]{Y. Tsipolitis}
\author[inst4,inst5]{S. Tzamarias}
\author[inst24]{A. Utrobicic}
\author[inst7,inst8]{I. Vai}
\author[inst1]{R. Veenhof}
\author[inst1]{L. Viezzi}
\author[inst7,inst8]{P. Vitulo}
\author[inst1,inst25]{C. Volpato}
\author[inst18]{X. Wang}
\author[inst10]{A. Weisenberger}
\author[inst1,inst26]{S. White}
\author[inst18]{Z. Zhang}
\author[inst18]{Y. Zhou}
\address[inst10]{Thomas Jefferson National Accelerator Facility, 12000 Jefferson Avenue, Newport News, VA 23606, USA}
\address[inst1]{European Organization for Nuclear Research (CERN), 1211 Geneve 23, Switzerland}
\address[inst2]{Université Paris-Saclay, F-91191 Gif-sur-Yvette, France}
\address[inst3]{IRFU, CEA, Université Paris-Saclay, F-91191 Gif-sur-Yvette, France}
\address[inst4]{Department of Physics, Aristotle University of Thessaloniki, University Campus, GR-54124, Thessaloniki, Greece}
\address[inst5]{Center for Interdisciplinary Research and Innovation (CIRI-AUTH), Thessaloniki 57001, Greece}
\address[inst6]{Department for Medical Physics, Ludwig Maximilian University of Munich, Am Coulombwall 1, 85748 Garching, Germany}
\address[inst7]{Dipartimento di Fisica, Università di Pavia, Via Bassi 6, 27100 Pavia, Italy}
\address[inst8]{INFN Sezione di Pavia, Via Bassi 6, 27100 Pavia, Italy}
\address[inst9]{Department of Physics and Astronomy, Stony Brook University, Stony Brook, NY 11794-3800, USA}
\address[inst11]{Institute of Nuclear and Particle Physics, NCSR Demokritos, GR-15341 Agia Paraskevi, Attiki, Greece}
\address[inst12]{Helmholtz-Institut für Strahlen- und Kernphysik, University of Bonn, Nußallee 14–16, 53115 Bonn, Germany}
\address[inst13]{Laboratório de Instrumentacão e Física Experimental de Partículas, Lisbon, Portugal}
\address[inst14]{Helsinki Institute of Physics, University of Helsinki, FI-00014 Helsinki, Finland}
\address[inst16]{University of Zagreb, Faculty of Electrical Engineering and Computing, 10000 Zagreb, Croatia}
\address[inst17]{European Spallation Source (ESS), Partikelgatan 2, 224 84 Lund, Sweden}
\address[inst18]{State Key Laboratory of Particle Detection and Electronics, University of Science and Technology of China, Hefei 230026, China}
\address[inst19]{Physikalisches Institut, University of Bonn, Nußallee 12, 53115 Bonn, Germany}
\address[inst20]{Friedrich-Alexander-Universität Erlangen-Nürnberg, Schloßplatz 4, 91054 Erlangen, Germany}
\address[inst21]{CEA-LIST, Diamond Sensors Laboratory, CEA Saclay, F-91191 Gif-sur-Yvette, France}
\address[inst22]{Queen’s University, Kingston, Ontario, Canada}
\address[inst23]{National Technical University of Athens, Athens, Greece}
\address[inst24]{Ruđer Bošković Institute, Bijenička cesta 54., 10 000 Zagreb, Croatia}
\address[inst25]{Department of Physics and Astronomy, University of Florence, Via Giovanni Sansone 1, 50019 Sesto Fiorentino, Italy}
\address[inst26]{University of Virginia, USA}
\fntext[fn1]{Now at Gran Sasso Science Institute, Viale F. Crispi, 7 67100 L'Aquila, Italy.}
\fntext[fn2]{Now at SOLEIL Synchrotron, L’Orme des Merisiers, Départementale 128, 91190 Saint-Aubin, France.}
\fntext[fn3]{Now at Department of Particle Physics and Astronomy, Weizmann Institute of Science, Hrzl st. 234, Rehovot, 7610001, Israel.}
\fntext[fn4]{Now at TÜV NORD EnSys GmbH \& Co. KG.}

%% file: 01_Introduction.tex
\section{Introduction}
The concept of picosecond timing detectors with a micro-pattern gaseous detector (MPGD) amplification structure was introduced by the PICOSEC collaboration with the development of PICOSEC Micromegas (PICOSEC-MM)~\cite{mm-picosec-bortfeldt,utrobicic-large-proto,mm-picosec-lisowska} technology. A PICOSEC-MM combines a Cherenkov radiator coated with a semi-transparent photocathode with a Micromegas~\cite{GIOMATARIS:199629} amplification structure to provide precision timing capabilities below 20 ps timing resolution. Potential  applications of this technology include large area and cost-effective time-of-flight (TOF) detectors for charged particles or single photon detectors for the readout of Cherenkov detectors for particle identification (PID) in future nuclear physics (NP) and high energy physics (HEP) experiments.   
In a PICOSEC-MM detector, a strong electric field in a narrow gas volume with a  $\sim$100~to~200~$\muup$m gap between the photocathode layer and the Micromegas structure is required for pre-amplification of the photoelectrons produced by the photocathode. The Micromegas amplification structure provides a second stage of amplification to achieve an overall gain greater than 10$^7$. The small gaps in the drift region of the PICOSEC-MM detector as well as those between the mesh and the anode readout PCB require a rigid PCB material to achieve excellent flatness ($\leq$~10~$\muup$m) and ensure a uniform timing response across the detector sensitive area.  
These stringent requirements pose a number of technical challenges for large size detectors, particularly with the mechanical stress caused by the stretching of the  Micromegas metallic mesh which can cause a significant bending of the PICOSEC anode  PCB~\cite{utrobicic-large-proto,Utrobicic_2023} and a severe degradation of the timing resolution. To reduce bending of the MM anode PCB, a hybrid board with a 4~mm thick embedded ceramic core material sandwiched between 2~$\times$~0.3~mm thick FR4 layers was developed and used for large (100~mm~$\times$~100~mm)  PICOSEC-MM prototypes. The hybrid PCB approach allows for the reduction of the maximum PCB deformation from 100~$\muup$m observed with the standard 3~mm-thick FR4 to less than 6~$\muup$m. For larger detectors (200~mm~$\times$~200~mm), an even thicker ceramic core would be required to keep the deformation at a minimum, resulting not only in an increase in detector thickness and cost but also in an increase in the material budget that could be detrimental to the performance of large area devices in future NP or HEP experiments. 
\\
In this paper we introduce the $\muup$RWELL-PICOSEC detector~\cite{urwellpicosec}, based on the same concept as the PICOSEC-MM detector but with $\muup$RWELL amplification  structure~\cite{Bencivenni:2015} replacing the Micromegas. Some of the challenges specific to the PICOSEC-MM technology, such as the stretching of the mesh or the presence of an array of supporting pillars in the detector active area, disappear with the $\muup$RWELL-PICOSEC technology, because unlike for PICOSEC-MM, the $\muup$RWELL foil used for the gas amplification is directly glued to the readout PCB and thus does not require mechanical stretching or supporting pillars to maintain a uniform gap. The gap is \textit{de facto} defined by the 50~$\muup$m thickness of the $\muup$RWELL foil. In section~\ref{sec:urwell}, we briefly introduce the $\muup$RWELL-PICOSEC concept with a detailed description of the single-pad  $\muup$RWELL-PICOSEC prototype design developed for the current study. In section~\ref{sec:testbeam} we describe the experimental setup used for the characterization in the high energy muon beam at CERN of several $\muup$RWELL-PICOSEC prototypes with different characteristics. The most relevant results are discussed in section~\ref{sec:results}. A summary of the results and perspectives for future optimization of the technology is presented in section~\ref{sec:conclusion}.

%% file: 02_small_proto.tex
\section{\texorpdfstring{$\muup$}{}RWELL-PICOSEC detector}
\label{sec:urwell}
%
%
\subsection{The \texorpdfstring{$\mu$}{}RWELL-PICOSEC detector concept}
\label{sec:concept}
\begin{figure}[!ht]
\centering
\includegraphics[width=0.995\columnwidth,trim={0pt 0mm 0pt 50mm},clip]{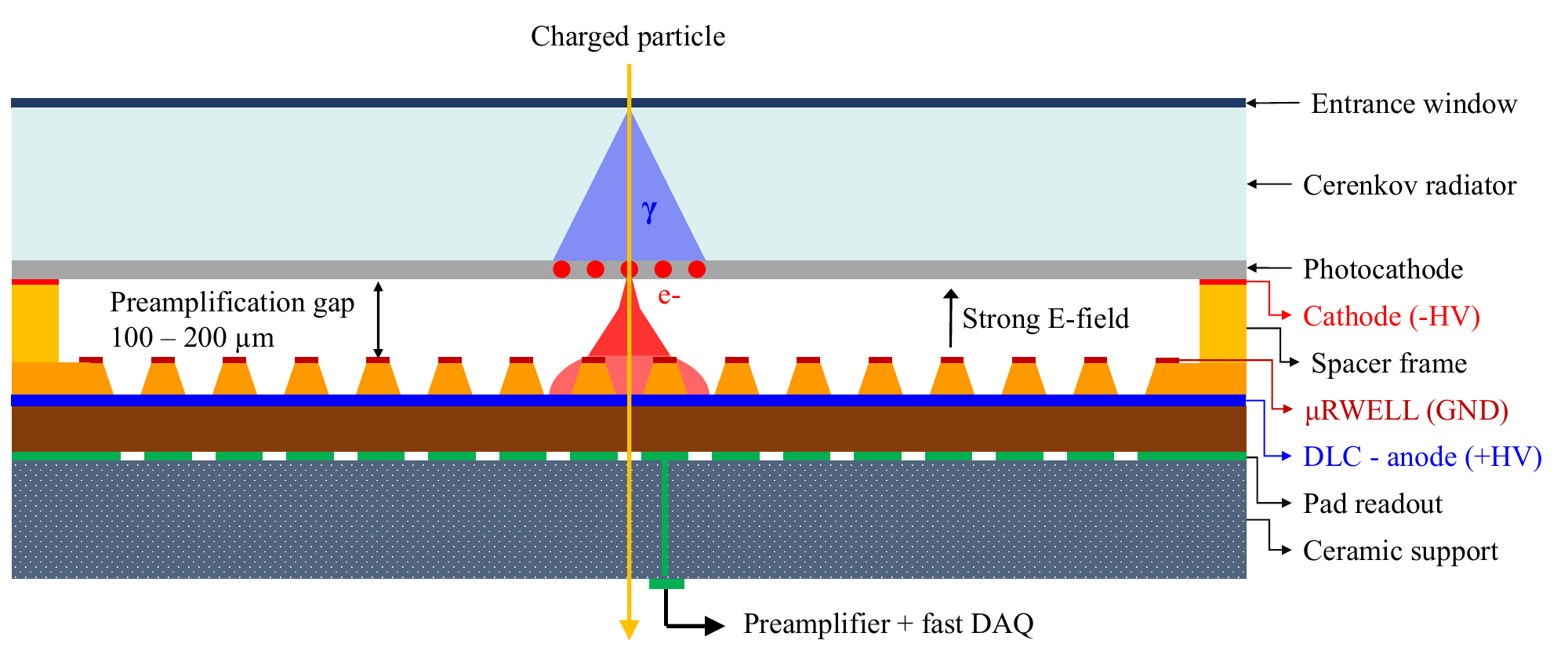}
\caption{\label{fig:pico_concept} Cross section of a $\muup$RWELL-PICOSEC detector layout (not to scale).}
\end{figure}

In a $\muup$RWELL-PICOSEC detector, the $\sim$3~mm thick ionization gas volume and cathode plane of a standard $\mu$RWELL detector are replaced by a Cherenkov radiator, typically a $\sim$3~mm thick Magnesium Fluoride  (MgF$_2$) crystal, with excellent light transmission in the UV region down to a wavelength of 115 nm. The inner surface of the radiator is coated with a thin layer of semi-transparent photocathode material such as Cesium Iodide (CsI) or Diamond Like Carbon (DLC). A high-energy charged particle traversing the detector produces Cherenkov light (photons) in the radiator that convert into photoelectrons in the thin photocathode layer. The resulting photoelectrons are pre-amplified in a strong electric field of approximately 20 to 40 kV / cm. This occurs in a narrow gas volume where an appropriate voltage is applied between the photocathode layer and the top electrode of the $\muup$RWELL PCB which serves as the anode, as illustrated in Fig.~\ref{fig:pico_concept}. The resulting charge is then multiplied a second time inside the $\muup$RWELL amplification structure to reach a gas gain approaching 10$^7$.
The electronic signal is received on the readout pad of the multilayer $\muup$RWELL PCB and amplified by a low capacitance fast preamplifier before being digitized by a high performance digitizer data acquisition system. The combination of photoelectrons generated from prompt Cherenkov photons in the highly localized region of the photocathode, two stages of electron amplification in the thin gas volumes subjected to a high electric field, and a fast readout electronics system enables the precise timing capabilities of a $\muup$RWELL-PICOSEC detector with sub 25 ps timing resolution.
%
%
\subsection{Single-pad \texorpdfstring{$\mu$}{}RWELL-PICOSEC prototypes}
\label{sec:assbly}
Small size single-pad $\muup$RWELL-PICOSEC prototypes were designed and developed to validate the  detector concept and investigate the geometrical parameters of the $\muup$RWELL amplification structure for the optimization of the timing performance.
\begin{figure}[!ht]
\centering
\includegraphics[width=0.995\columnwidth,trim={0pt 0mm 0pt 10mm},clip]{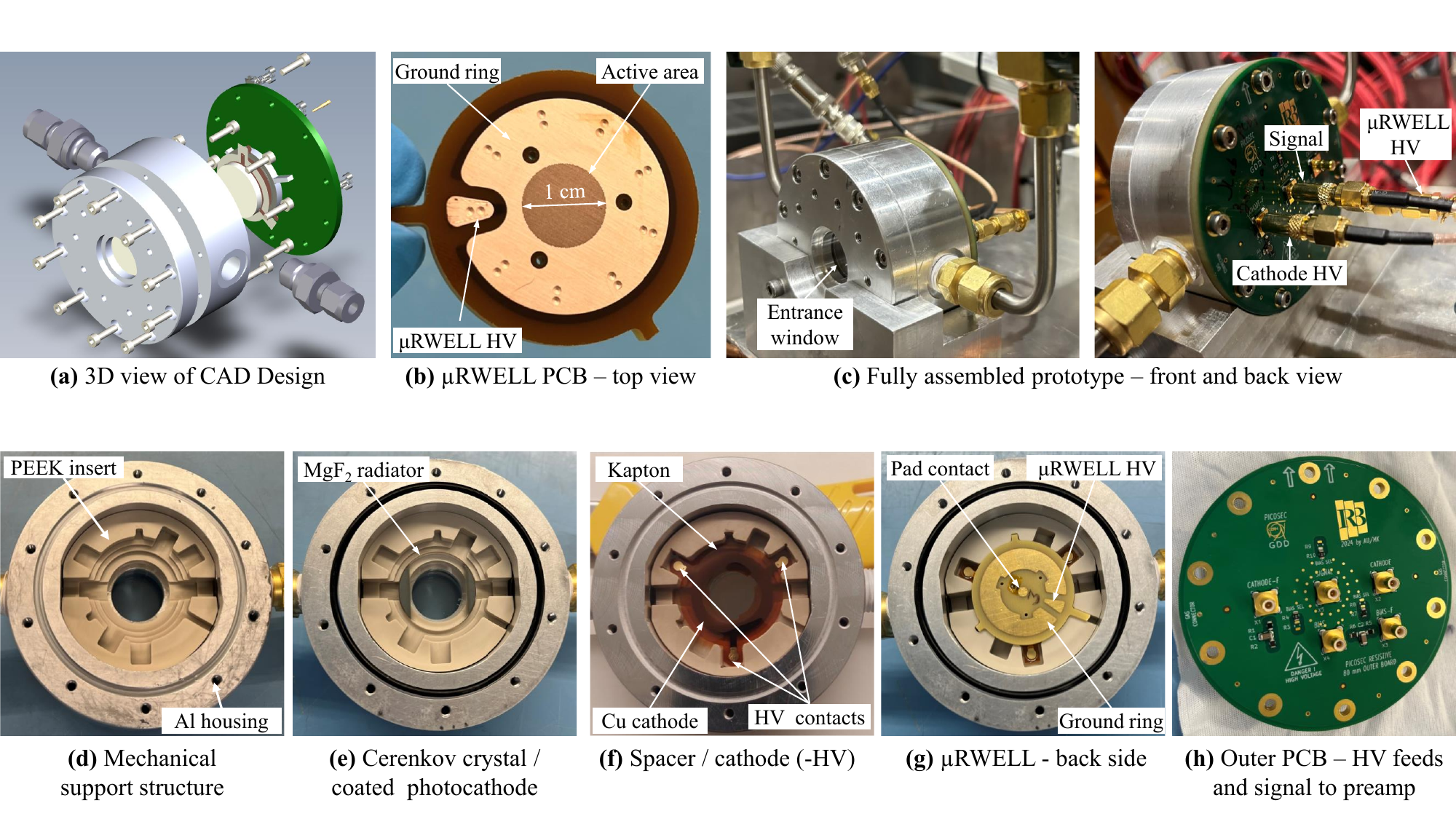}
\caption{\label{fig:assembly} Top panel -- \textbf{(a):} Exploded 3D view of the single-channel $\muup$RWELL-PICOSEC prototype design; \textbf{(b):}  Picture of the $\muup$RWELL PCB, a key component of the detector; \textbf{(c):} Front and rear view of the fully assembled prototype; Bottom panel -- \textbf{(d):} to \textbf{(g):}  Assembly steps in sequential order of a single-pad $\muup$RWELL-PICOSEC detector;  \textbf{(h):}  The outer PCB closes and seals the aluminum housing and provides a HV supply to the pre-amplifier and a robust grounding scheme.}
\end{figure}
The 3D CAD view of a single-pad $\muup$RWELL-PICOSEC prototype design is in panel \textbf{(a)} of Fig.~\ref{fig:assembly} and panel \textbf{(b)} is a photograph of the detector anode PCB composed of the $\muup$RWELL amplification layer glued to the pad readout PCB. Photographs of the front and back sides of the fully assembled detector in a test stand are in panel \textbf{(c)}. The pictures in the bottom panels show the different assembly steps of the prototype in the mechanical support structure consisting of the aluminum housing and the polyether ether ketone (PEEK) insert. The step-like inner structure of the PEEK insert matches in diameter and thickness the different parts of the assembly stack which start with the insertion of the MgF$_2$ radiator  shown in panel \textbf{(e)}. A 150~$\muup$m thick Kapton spacer with a 17~$\muup$m thick Cu electrode ring on one side shown in panel  \textbf{(f)} is laid on the radiator and defines a 167~$\muup$m pre-amplification gap between the photocathode and the anode PCB. The Cu cathode ring directly in contact  with the photocathode sets the voltage for the strong electric field in the drift region. The $\muup$RWELL PCB is positioned upside-down as shown in panel \textbf{(g)} with the back side showing the electrical contact of the pad at the center surrounded by the detector ground ring and the electrode supplying the HV to the $\muup$RWELL DLC anode. The outer PCB shown in panel \textbf{(h)} is used to close the aluminum housing with an O-ring, a set of screws and a gasket for gas sealing. This outer board is used to distribute the voltage to different elements of the detector and conduct the signal from the detector readout pad to the fast preamplifier of the readout and DAQ chain. The design of the mechanical support structure of the  single-pad $\muup$RWELL-PICOSEC prototype was inspired from PICOSEC-MM prototypes~\cite{utrobicic-small-proto}, with a few minor modifications on the outer board to adapt the HV powering scheme to the resistive $\muup$RWELL. The design is optimized to facilitate the assembly of single-pad PICOSEC prototypes, allowing rapid replacement of detector components such as the Cherenkov radiator, the spacer ring, and the $\muup$RWELL PCB. The full assembly of a prototype can be completed in under five minutes in a clean room which is critical for minimizing the exposure of the CsI photocathode when replacing $\muup$RWELL PCBs or Kapton spacers for lab studies or test beams.

%% file: 03_testbeam_setup.tex
\section{Development of single-pad \texorpdfstring{$\muup$}{}RWELL-PICOSEC prototypes}
\label{sec:testbeam} 
%
%
\subsection{Design of \texorpdfstring{$\mu$}{}RWELL PCBs with different geometrical parameters}
\label{subsec:parameters}
 Several single-pad $\muup$RWELL-PICOSEC prototypes with different $\muup$RWELL hole geometries and readout patterns were fabricated and characterized in a high energy muon beam at CERN.
\begin{figure}[!ht]
\centering
\includegraphics[width=0.995\columnwidth,trim={0pt 0mm 0pt 0mm},clip]{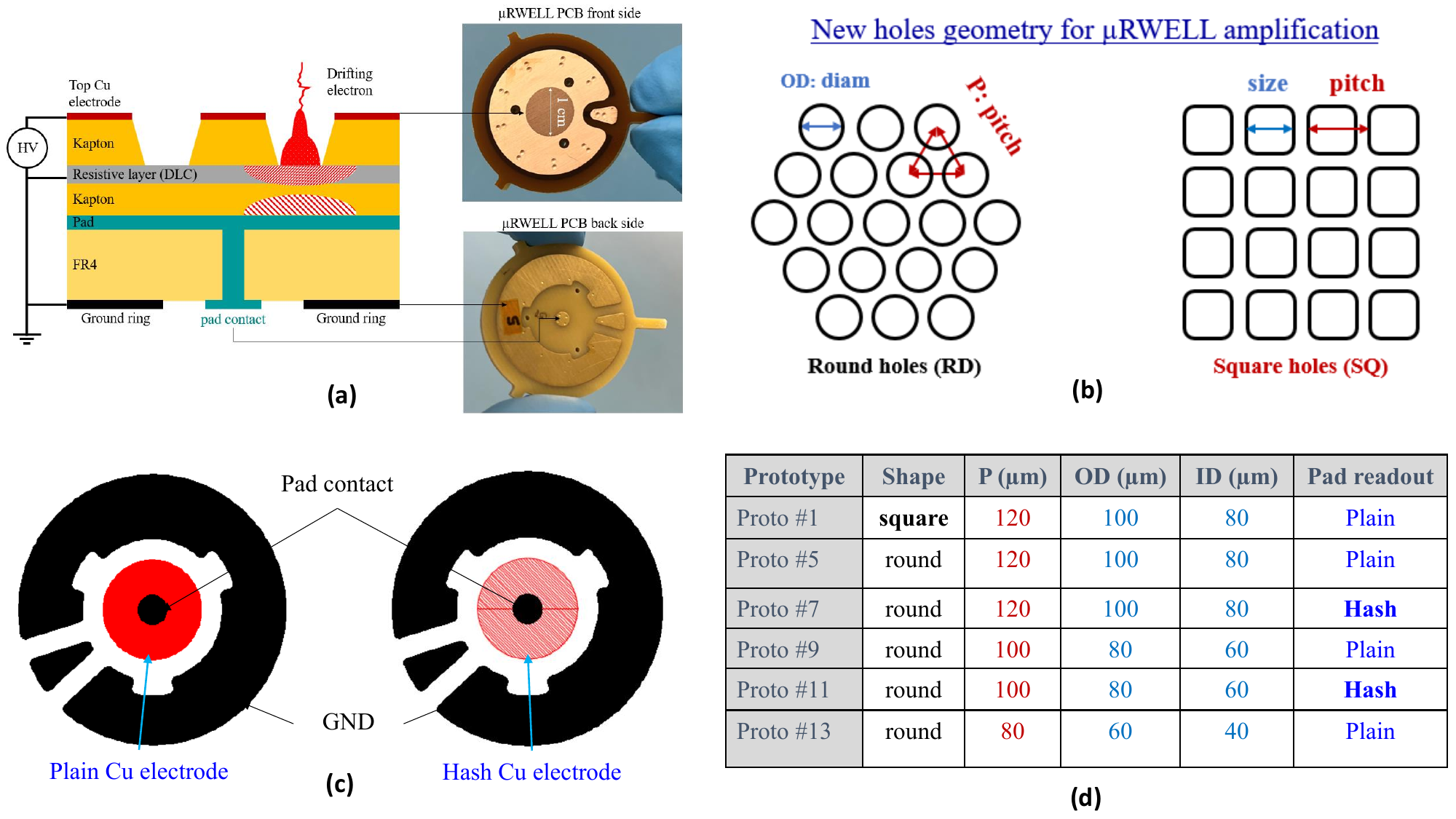}
\caption{\label{fig:param} \textbf{(a)}: Layout and pictures of the $\muup$RWELL PCBs;  \textbf{(b)}: Cartoon of two  $\muup$RWELL hole geometries,  round holes (left) and square holes (right);  \textbf{(c)}: Cartoon of two readout pad geometries,  plain Cu electrode (left)   and hashed geometry Cu electrode to minimize detector capacitance  (right); \textbf{(d)}: Table summarizing the geometrical parameters of the  $\muup$RWELL PCBs for different prototypes.}
\end{figure}
Fig.~\ref{fig:param} panel \textbf{(a)} is a cross section diagram of the $\muup$RWELL PCB with the pad readout design and photographs of the front and back of the PCB. Panel \textbf{(b)} are sketches of two different $\muup$RWELL hole geometries. The one on the left has round holes with a hexagonal arrangement pattern of a standard $\muup$RWELL, while the one on the right has a square hole pattern. The goal is to study the impact of the hole shape on the electric field lines and on the path of the charges drifting toward the readout pad. In panel \textbf{(c)} are sketches of two readout pad patterns, one with a plain Cu pick-up electrode and one with a pattern we call hashed Cu; it reduces the total area of the Cu electrode by half to minimize the detector input capacitance.  Finally, panel \textbf{(d)} is a table of the parameters of the different $\muup$RWELL PCBs tested in the beam.
%
%
%
\subsection{CERN test beam setup}
\label{subec:setup} 
A tracking and timing telescope was fabricated at JLab for the characterization of single-pad $\muup$RWELL-PICOSEC prototypes in test beams. 
\begin{figure}[!ht]
\centering
\includegraphics[width=0.995\columnwidth,trim={0pt 0mm 0pt 0mm},clip]{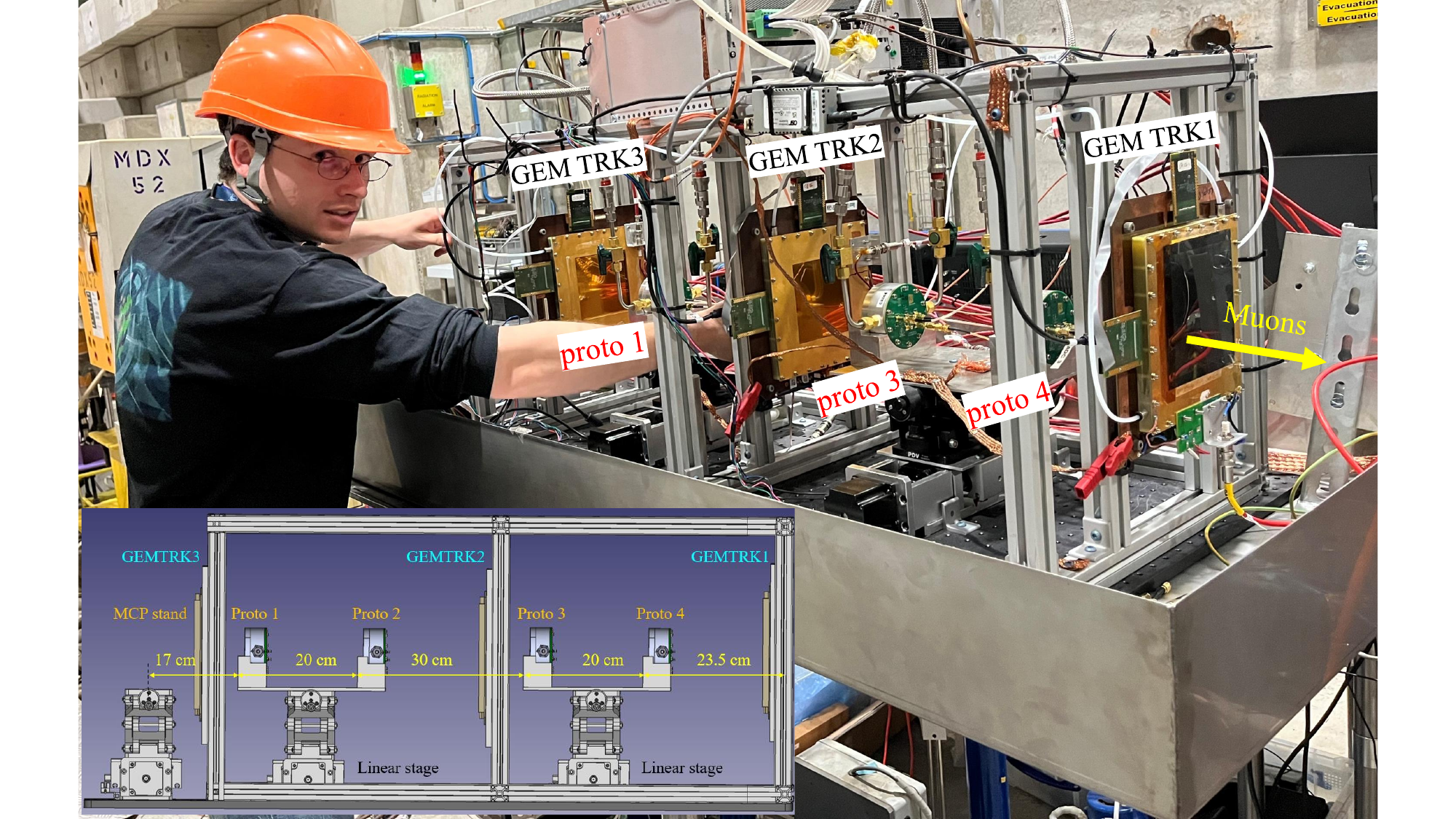}
\caption{\label{fig:telescope} $\muup$RWELL-PICOSEC telescope with three GEM trackers, a MCP-PMT detector for reference time and a stand for four single-pad $\muup$RWELL-PICOSEC prototypes. A schematic of the telescope layout is shown on the bottom left.}
\end{figure}
 The telescope is composed of three small (10 cm $\times$ 10 cm) triple-GEM detectors with X-Y strip readouts that provide reference tracks and one Hamamatsu R3809U--50 MCP-PMT~\cite{mcppmt} detector, installed upstream of the first GEM tracker that acts as a trigger and provides a precise reference time. Two platforms, mounted on a complex set of horizontal (X) and vertical (Y) linear stages between the three GEM trackers, are used as test stands for  four  single-pad $\muup$RWELL-PICOSEC prototypes. The X-Y linear stages are for the alignment of the active area of the prototypes being tested with respect to the MCP-PMT timing detector. A picture of the  $\muup$RWELL-PICOSEC telescope is shown in Fig.~\ref{fig:telescope} along with the schematic of the layout in the bottom left corner. The "COMPASS" gas mixture Ne:CF$_4$:C$_2$H$_6$ (80:10:10) is used as the fast gas for the PICOSEC prototypes. The standard Ar:CO$_2$ (80:20) gas mixture is used for the GEM trackers. \\
 The signals from the readout pads of the $\muup$RWELL-PICOSEC prototypes are amplified with a custom-developed RF pulse amplifier card~\cite{preamp_Hoarau_2021} optimized for PICOSEC with built-in discharge protection up to 350 V, 650 MHz bandwidth, 38 dB gain and 75 mW power consumption. The amplified signal is digitized  with a high performance LECROY WR8104 oscilloscope~\cite{oscillo} that operates at a 1.0 GHz analogue bandwidth at a sampling rate of 10 GSamples/s. Two such oscilloscopes are needed for all four prototypes installed on the telescope. Ch:1 and Ch:2 of each oscilloscope are connected to two channels of the amplifier card from two prototypes. The MCP-PMT signal is split by a 50$\omegaup$ splitter and one output is sent to Ch:3 of the oscilloscope to be digitized. The GEM trackers are read out with APV25-based SRS electronics~\cite{apv25,srs2011}. DATE and amoreSRS~\cite{aliceSoft, amoresrs} software are used for data acquisition (DAQ) and data online monitoring. The SRS crate is triggered by the MCP-PMT signal. To synchronize the SRS events and the fast signals sent to the oscilloscope DAQ, the NIM signal at the output of the SRS trigger is sent to the external trigger channel of the two oscilloscopes and the triggered event number (event ID), recorded in the SRS data stream, is simultaneously sent as a bit stream to Ch:4 of the oscilloscopes. The SRS data are decoded with amoreSRS, which is also used for the offline analysis of the GEM trackers' data. The telescope was commissioned in beam at CERN during the July 2024 test beam campaign and fully operated in the August 2024 test beam for the study of several single-pad $\muup$RWELL-PICOSEC prototypes. 
%
%
%
\subsection{\texorpdfstring{$\mu$}{}RWELL-PICOSEC signal processing}
\label{subsec:signal} 
The MCP-PMT signal is used as the reference time for the analysis to extract the precision timing information from the $\muup$RWELL-PICOSEC data. The MCP-PMT has an intrinsic time resolution on the order of 5~ps~\cite{mcp_timing} which is uniform in a 4-mm radius around the center. The resolution is slightly degraded to about 8~ps when the MCP-PMT signal is split in two as described in section ~\ref{subec:setup}. This split has a negligible impact on the time resolution of the $\muup$RWELL-PICOSEC prototypes under test.
\begin{figure}[!ht]
\centering
\includegraphics[width=0.995\columnwidth,trim={0pt 0mm 0pt 90mm},clip]{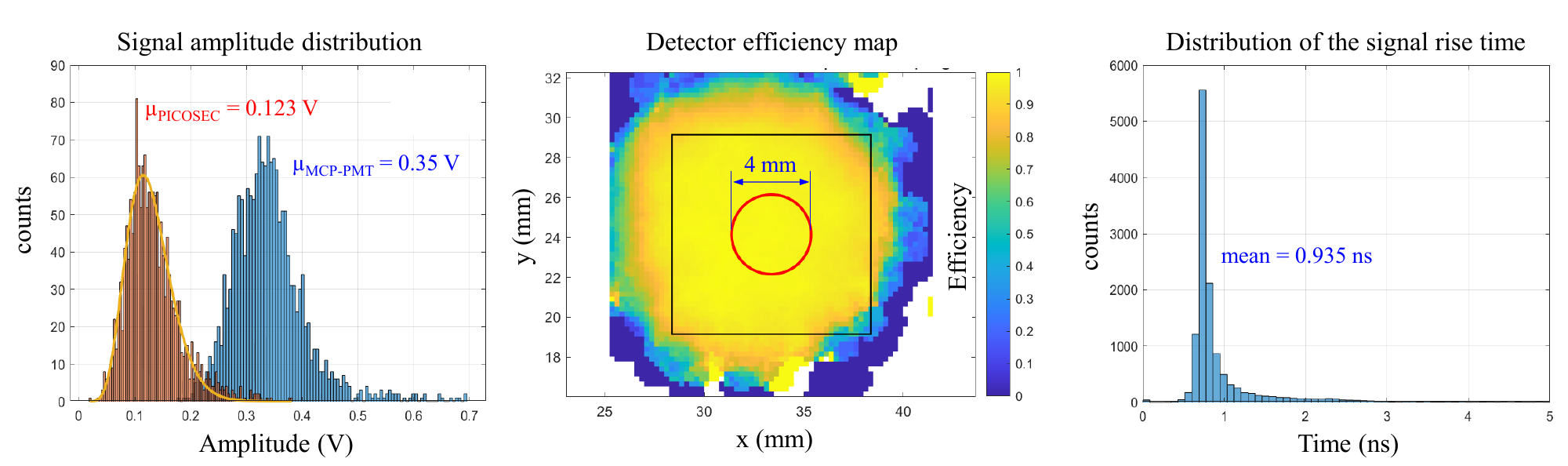}
\caption{\label{fig:eff} \textbf{(Left)}: Distributions of the signal amplitude for the MCP-PMT and one $\muup$RWELL-PICOSEC proto~\#5; \textbf{(center)} 2D mapping of the spatial distribution of the detector efficiency where the area inside the red circle represents the projection of the 4~mm diameter acceptance region of the MCP-PMT with minimal and uniform time resolution; \textbf{(right)} Distribution of the rise time of the $\muup$RWELL-PICOSEC proto~\#5.
}
\end{figure}
The  panel on the left of Fig.~\ref{fig:eff} shows the distribution of signal amplitudes for both the MCP-PMT (blue) and the $\muup$RWELL-PICOSEC proto~\#5 (brown) operating with a CsI photocathode at 440~V and a $\muup$RWELL DLC anode at 265~V. The equivalent electric fields are $\sim$26~kV~/~cm in the pre-amplification drift region and 53~kV~/~cm in the $\muup$RWELL amplification holes.  The 2D efficiency map of $\muup$RWELL-PICOSEC proto~\#5 is shown in the center of Fig.~\ref{fig:eff}. Only the events with hits inside the 4-mm radius red circle are considered in the analysis. In this region, the efficiency of proto~\#5 reaches 98.5\% for these voltage settings.
\begin{figure}[!ht]
\centering
\includegraphics[width=0.995\columnwidth,trim={0pt 0mm 0pt 0mm},clip]{fig_picosec_signal.pdf}
\caption{\label{fig:signal} \textbf{(Top)}: Raw waveform signal recorded in the oscilloscope of the reference MCP-PMT signal (left) and the $\muup$RWELL-PICOSEC signal (right). The latter is composed of the fast component of the induced electron signal and the slow component of the ion tail; \textbf{(bottom)}: Zoomed view of the leading edge of the two signals fitted to a sigmoid function for the MCP-PMT (left) and $\muup$RWELL-PICOSEC (right).}
\end{figure}
The typical waveform signals of the $\muup$RWELL-PICOSEC prototype and the MCP-PMT detector recorded in the oscilloscopes are shown in the top panels of Fig.~\ref{fig:signal}. As expected, the raw signal from the MCP-PMT has a very sharp peak with a width of a few hundred picoseconds. The raw signal of the $\muup$RWELL-PICOSEC detector has two components, a fast signal with a steep rise time and sharp peak from the induced signal of the electron drift in the amplification regions and a slow component from the ion back flow. An example of the distribution of the rise time of the fast electron peak of the $\muup$RWELL-PICOSEC prototype is shown in the right panel of Fig.~\ref{fig:eff} with a mean value of 0.935~ns. 
\begin{equation}
V(t) = \frac{P_{0}}{1+exp(-P_{2}\times (t-P_{1}))} + P_{3} \label{eq:sigmoid}
\end{equation}
To extract the timing information of the $\muup$RWELL-PICOSEC, the leading edges of the signal pulses of the MCP-PMT and the prototypes are fitted to a sigmoid function~\eqref{eq:sigmoid}  where $P_{0}$ and $P_{3}$ represent the maximum and minimum values of the signal respectively. $P_{1}$ corresponds to the inflection point where the slope of the curve changes sign and $P_{2}$ characterizes the steepness of the sigmoid change, directly related to the signal rise time. An example of sigmoid fits is shown in the bottom plots of Fig.~\ref{fig:signal} for the MCP-PMT (left) and the $\muup$RWELL-PICOSEC prototype (right).
The time stamps corresponding to the 20\% Constant Fraction (CF)~\cite{mm-picosec-bortfeldt} are computed based on~\eqref{eq:cftime} for both the MCP-PMT and the prototypes under test, where $V_{max}$ is the amplitude of the electron-peak. The sigmoid fit provides an essential tool for determining the precise timing position at the 20\% Constant Fraction (CF) level.
\begin{equation}
t_{ST} = P_{1}- \frac{1}{P_{2}} \log \left[\frac{P_{0}}{0.2\times V_{max} - P_{3}} - 1\right]
    \label{eq:cftime}
\end{equation}
\subsection{Signal Arrival Time (SAT) and Time Resolution}
\label{subsec:timeres} 
\begin{figure}[!ht]
\centering
\includegraphics[width=0.995\columnwidth,trim={0pt 0mm 0pt 45mm},clip]{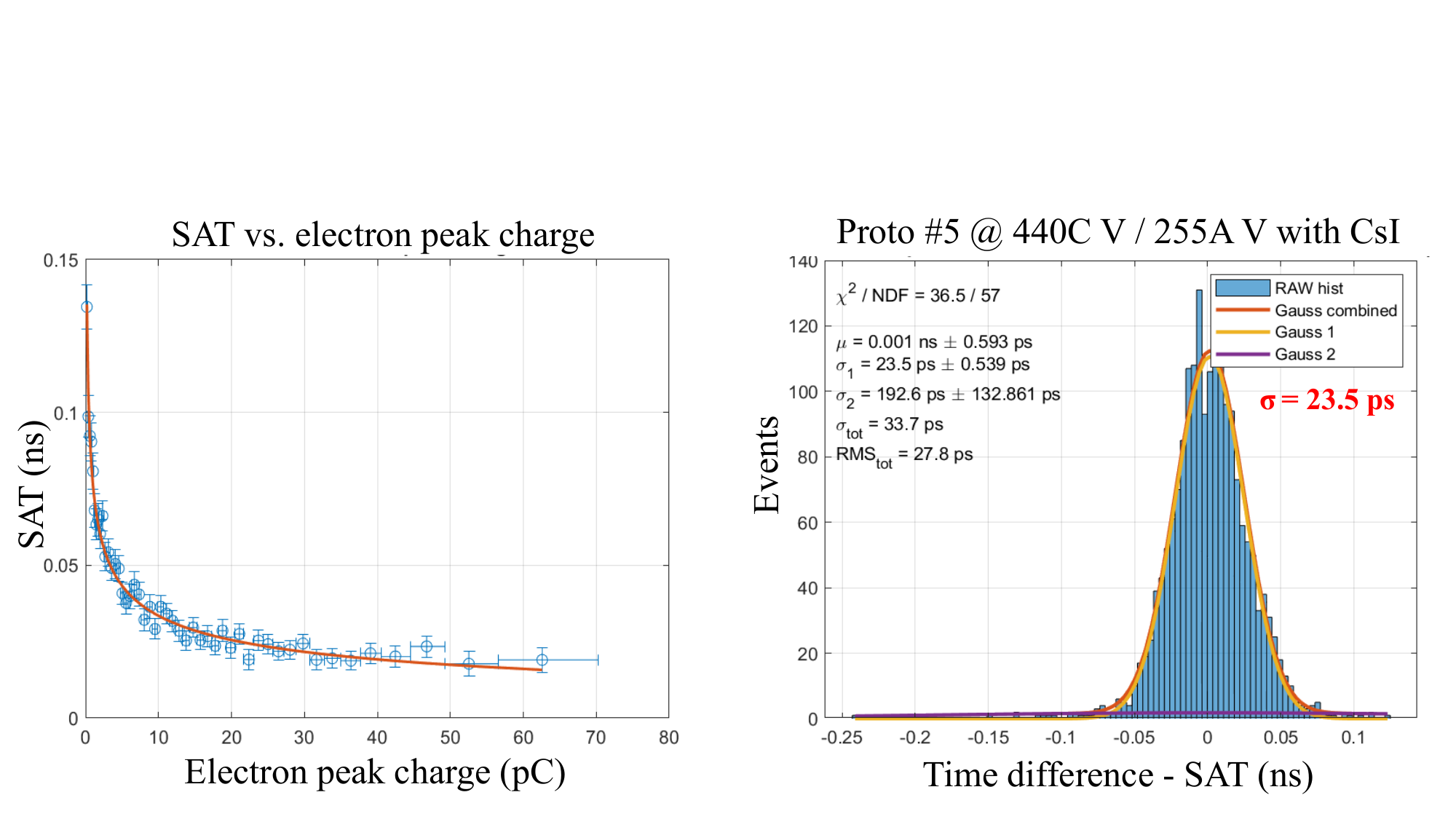}
\caption{\label{fig:sat} \textbf{(Left)}: SAT as a function of the electron peak charge. The red curve is a polynomial fit to the data; \textbf{(right)}: Distribution plots of the time difference between $\muup$RWELL-PICOSEC proto~\#5 with a CsI photocathode and the MCP-PMT reference detector.}
\end{figure}

The signal arrival time (SAT) is defined as the difference between the $\muup$RWELL-PICOSEC prototype time and the MCP-PMT time. The SAT is calculated for each event and plotted as a function of the electron peak (epeak) charge in pC of the $\muup$RWELL-PICOSEC prototype as shown on the left of Fig.~\ref{fig:sat}. The data points in the graph represent the mean value of the time differences grouped in bins for a given epeak charge and plotted against the epeak charge. The SAT data are fitted to a polynomial function (red curve in Fig.~\ref{fig:sat}), used as a calibration function when computing the time resolution of the $\muup$RWELL-PICOSEC prototype to correct for time jitter of the electron peak with the signal amplitude. A geometric cut is also applied to select data within a 2 mm radius around the center of the prototypes as shown with the red circle in the 2D map of the center panel of Fig.~\ref{fig:eff}. This cut guarantees that only the events at the center of the MCP-PMT where the time resolution is the best are analyzed. In this case, the contribution of the time jitter of the reference detector to the time resolution of the prototypes under test is negligible. The time difference shown in the distribution plot on the right of Fig.~\ref{fig:sat} is corrected on an event-by-event basis by subtracting the SAT value for a given electron peak charge using the correction function of Fig.~\ref{fig:sat}. The distribution is then fitted to a double Gaussian function where the width $\sigmaup$ of the first Gaussian fit is used to define the time resolution of the prototype. A time resolution of 23.5~$\pm$~0.53 ps was achieved with proto~\#5 equipped with a 18~nm CsI photocathode at 440~V and the $\muup$RWELL anode at 255~V. This represents the best performance achieved for all six $\muup$RWELL-PICOSEC prototypes tested in the two 2024 test beam campaigns.

%% file: 04_results.tex
\section{Results \& discussion}
\label{sec:results} 
%
%
\subsection{Characterization  of \texorpdfstring{$\mu$}{}RWELL-PICOSEC proto~\#5
in the April 2024 test beam}
\label{subsec:proto5_hvscan} 
A $\muup$RWELL-PICOSEC prototype (proto~\#5) was first tested during the April 2024 CERN RD51 test beam campaign in the MM-PICOSEC telescope while the $\muup$RWELL-PICOSEC telescope was commissioned for use in the July 2024 test beam. The MM-PICOSEC telescope had the same layout configuration as the one described in Fig.~\ref{fig:telescope}. A high voltage scan of the cathode and anode was performed on the prototype in the 150~GeV~/~c muon beam to evaluate the detector performance and study the impact of both pre-amplification and amplification on the detector timing response.
\begin{figure}[!ht]
\centering
\includegraphics[width=1.\columnwidth,trim={0pt 0mm 0pt 85mm},clip]{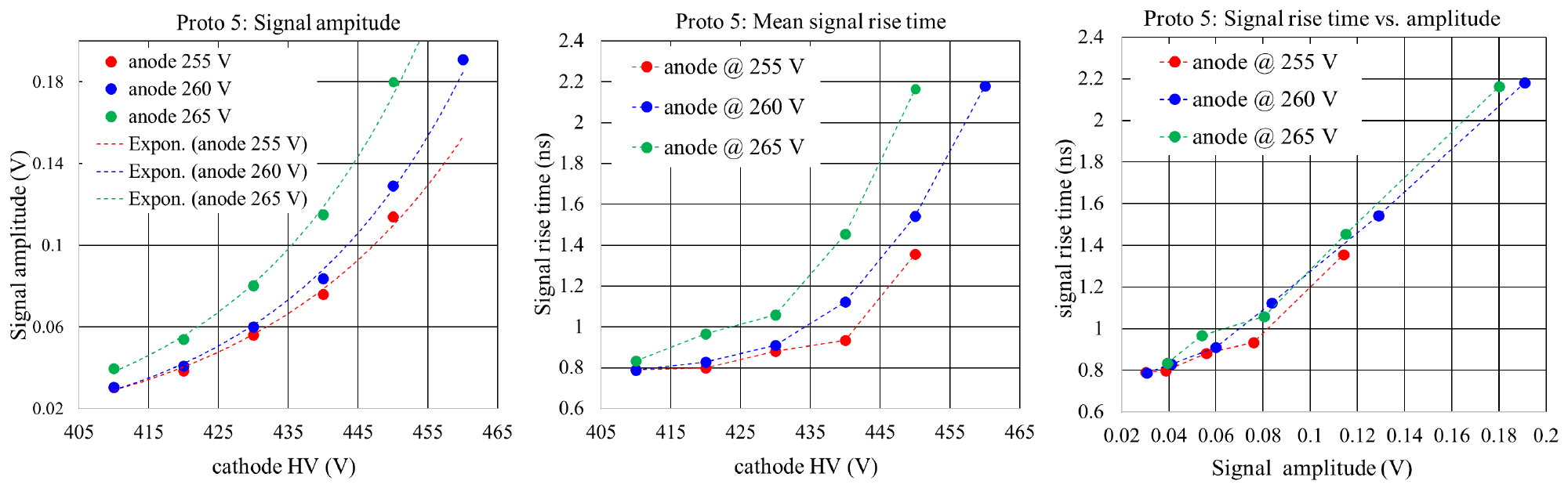}
\caption{\label{fig:perf} \textbf{(Left)}: Signal amplitude as a function of  cathode voltage; \textbf{(center)}: Signal rise time as a function of cathode voltage; \textbf{(right)}: Signal rise time as a function of  signal amplitude.}
\end{figure}
The plots in the left panel of Fig.~\ref{fig:perf} show the signal amplitudes of proto~\#5 as a function of the cathode voltage varying from  410~V to 460~V. The cathode voltage correspond to a drift field varying from 24.5 kV~/~cm to 27.5~kV~/~cm for three different DLC anode voltages, 250~V, 255~V and 260~V. The expected exponential shape (dotted lines) of the amplitude (i.e. detector gain) with the cathode voltage is observed for all three anode voltage settings. In the center panel of Fig.~\ref{fig:perf}, the plots show the mean value of the rise time of the detector signal for the same cathode and anode voltage scans. An example of the rise time distribution is shown in the right panel of Fig.~\ref{fig:eff}. The signal rise time also has an exponential-like dependence on the cathode HV and a quasi-linear dependence on the signal amplitude as shown in the plots in the right panel of Fig.~\ref{fig:perf}. The reason for the strong dependence of the rise time on the signal amplitude is still under investigation. Ideally, we would  expect the rise time to be independent of the amplitude.\\

\begin{figure}[!ht]
\centering
\includegraphics[width=1.\columnwidth,trim={0pt 0mm 0pt 85mm},clip]{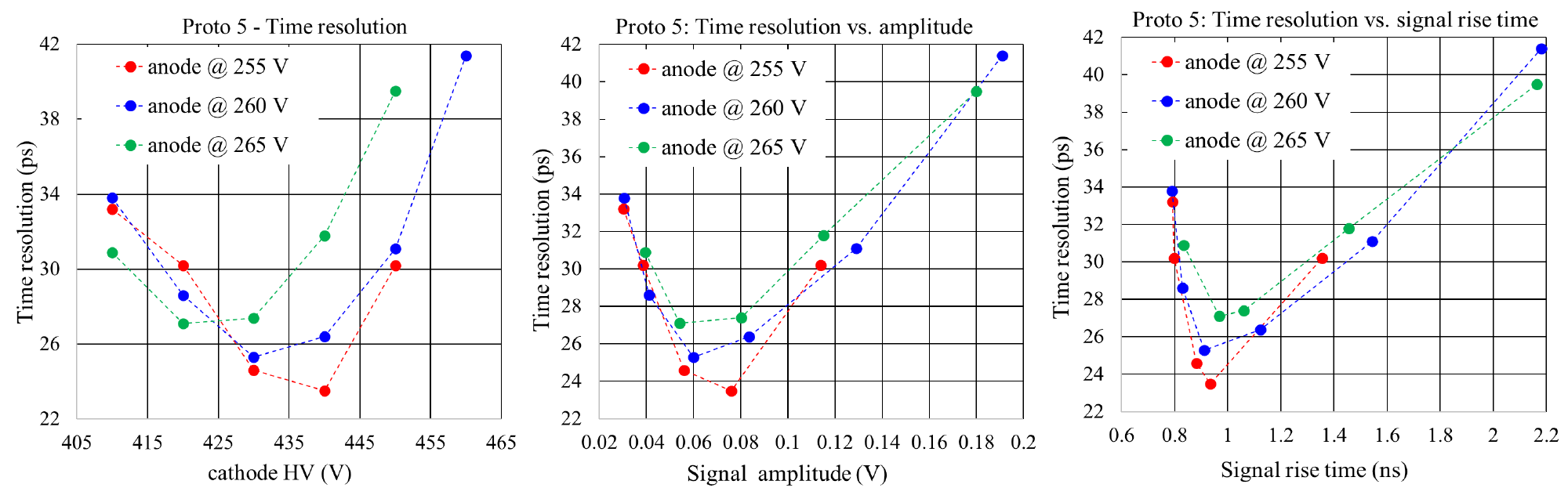}
\caption{\label{fig:pico5_timing} Time resolution of $\muup$RWELL-PICOSEC proto~\#5 as a function of - \textbf{(left)}: applied cathode HV (V); \textbf{(center)}: signal amplitude (mV); \textbf{(right)}: signal rise time (ns).}
\end{figure}

The dependence of the time resolution of  proto~\#5 on the cathode voltage resembles a bell curve for each anode voltage, as shown in the left panel of Fig.~\ref{fig:pico5_timing}. The minimum of the curve represents the optimal resolution achievable for the given anode voltage. The best time resolution performance achieved during the test beam campaign was 23.5~ps at a cathode voltage of 440~V and an anode voltage of 255~V. For the other two anode voltages, 260~V and 265~V, the best timing performance was 25.3~ps and 27.1~ps achieved at a cathode voltage of 430~V and 420~V respectively.
%
It is still under investigation as to why the time resolution of the $\muup$RWELL-PICOSEC detector reaches a minimum (best performance) at a given cathode voltage and then starts to increase (degrade) at larger voltages. The dependence of the time resolution of the $\muup$RWELL-PICOSEC detector on the cathode voltage is different from what is observed in literature~\cite{mm-picosec-bortfeldt}. Data from the MM-PICOSEC test beam show a steady improvement (decrease) up to the point where the detector becomes unstable in the strong electric field. The best time resolution of a MM-PICOSEC detector is generally reached at a voltage close to the instability regime of the detector. A possible explanation could be that at higher voltages, the strong electric field in the $\muup$RWELL holes causes strong field line distortions toward the holes which impact the path length of the charges drifting to the holes. Another possible explanation is that at higher gain, micro-discharges occur which do not affect the operational stability of the detector but deteriorate the quality of the electronic signal and subsequently the time resolution. An additional effect contributing to the degradation of the time resolution for higher gain could be attributed to the sampling resolution of the oscilloscope. This is however not an inherent limitation for the detector timing performance. All hypotheses are investigation and will be the focus of the future test beam campaign and simulation studies of the $\muup$RWELL-PICOSEC. The graphs in the center panel of Fig.~\ref{fig:pico5_timing} show the time resolution as a function of the signal amplitude. A minimum is reached for signal amplitudes between about 0.06~V and 0.08~V for each anode setting. The time resolution is plotted against the detector signal rise time in the right panel. 
%
%
\subsection{Characterization of prototypes with different geometrical parameters in the July 2024 test beam}
\label{subsec:proto_comapre} 
\begin{figure}[!ht]
\centering
\includegraphics[width=1.\columnwidth,trim={0pt 0mm 0pt 85mm},clip]{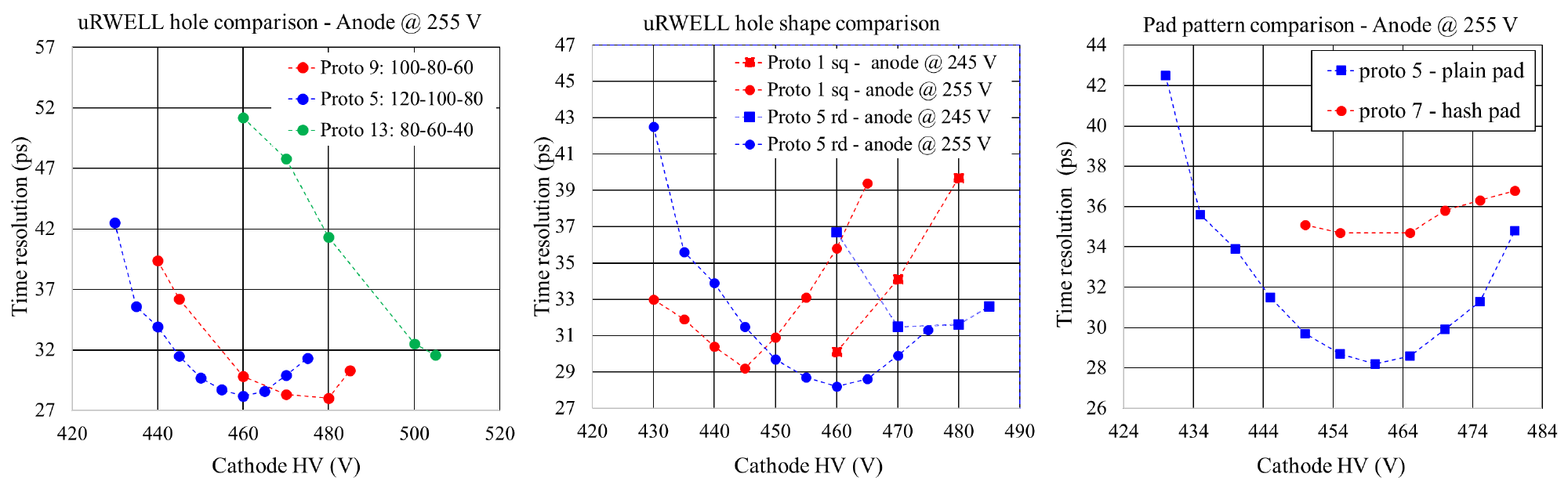}
\caption{\label{fig:compare} Time resolution as function of the cathode HV for - \textbf{(left)}: three prototypes with different $\muup$RWELL hole pitches and diameters; \textbf{(center)}: two prototypes, one with round $\muup$RWELL holes (proto~\#5) and one with square $\muup$RWELL holes (proto 1); \textbf{(right)}: two prototypes, one with a plain readout pad (proto~\#5) and one with a hashed readout pad (proto 7).}
\end{figure}

The single-pad $\muup$RWELL-PICOSEC prototypes listed in the table of Fig.~\ref{fig:param} were all tested during a second RD51 test beam campaign in July 2024 to study various $\muup$RWELL hole features such as the shape, diameter and pitch as well as the readout pad pattern for the optimization of the time resolution at voltage settings that guarantee detector robustness for operational stability. All prototypes were installed in the $\muup$RWELL-PICOSEC telescope setup of Fig.~\ref{fig:telescope} for  anode and cathode voltage scans. The timing resolution plots as a function of the cathode voltage are shown in the left panel of Fig.~\ref{fig:compare} for three prototypes with  different hole diameters and pitches but with a constant pitch-to-hole difference of 20~$\muup$m at a fixed anode voltage of 255~V. The details of the pitch, outer diameter (OD) and inner diameter (ID) of the holes are shown in the table of Fig.~\ref{fig:param}. The pitches of the $\muup$RWELL holes of the three tested prototypes \#5, \#9 and \#13 are 120, 100 and 80~$\muup$m respectively, and the outer diameters are 100, 80 and 60~$\muup$m resulting in a pitch-to-hole diameter of 0.833, 0.8 and 0.75. The best performance was 28~ps for  proto~\#9 at 480~V, 28.2~ps for proto~\#5 at  460~V and 31.6~ps for proto~\#13 at 505~V. The hole parameters of the $\muup$RWELL-PICOSEC prototypes differ significantly from those of a standard $\muup$RWELL which has a pitch, OD and ID of 140, 70 and 50~$\muup$m respectively and a significantly smaller pitch-to-diameter ratio of 0.5.  Preliminary studies of the $\muup$RWELL-PICOSEC prototype (proto~\#0) with standard hole parameters, shows a worse time resolution performance with the best achieved on the order of 45~ps. Proto~\#0 was from an earlier production batch and is not  listed in the table of ~\ref{fig:param}. In general, prototypes with large pitch-to-diameter ratios have better time resolution, although the absolute hole diameter is also an important parameter to optimize detector performance. 

The plots in the center panel of Fig.~\ref{fig:compare} show the comparison between two prototypes with the same $\muup$RWELL hole pitch of 120~$\muup$m but different hole shapes. Proto~\#5 (blue) has a round hole with an OD of 100~$\muup$m and proto~\#1 (red) has a square hole with a size of 100~$\muup$m. A cartoon illustration of the hole geometry for the two prototypes is shown in the top right quarter of Fig.~\ref{fig:param}. Cathode voltage scans were performed for two anode voltage settings, 245~V (square dots) and 255~V (round dots). The plots show better timing  performance of proto~\#5 with the round holes compared to  the square holes of proto~\#1, especially for the anode voltage at 255~V. For the anode voltage of 245~V, proto~\#1 shows better performance at the lower cathode voltage ($\le$ 470~V), although we did not have enough data points at the lower voltage to reach the minimum before the time resolution started increasing again. Additional measurements will be performed to reach a definitive conclusion on the impact of the hole shape on the time resolution optimization. In the right panel of Fig.~\ref{fig:compare}, the plots show the comparison of two prototypes with identical $\muup$RWELL hole parameters but with different readout pad patterns. The two prototypes have 1-cm  diameter pads, but proto~\#5 (blue) has a plain Cu electrode while proto~\#7 (red) has a hashed Cu electrode as described in the bottom left quarter of Fig.~\ref{fig:param}. The Cu area of the hashed pad of proto~\#7 is roughly half the size of the plain pad so we expected a minimal contribution of the detector input capacitance to the signal-to-noise ratio and consequently a better timing performance. That being said, the test beam results show far better timing performance with proto~\#5 with the plain Cu electrode with the best timing resolution of 28.2~ps (blue dots) than proto~\#7 with the hashed Cu electrode with the best timing resolution of 34.7~ps (red dots). 
%
%
\subsection{Time resolution of various prototypes with a CsI photocathode}
\label{subsec:csi_photoC} 
\begin{figure}[!ht]
\centering
\includegraphics[width=0.95\columnwidth,trim={0pt 0mm 0pt 0mm},clip]{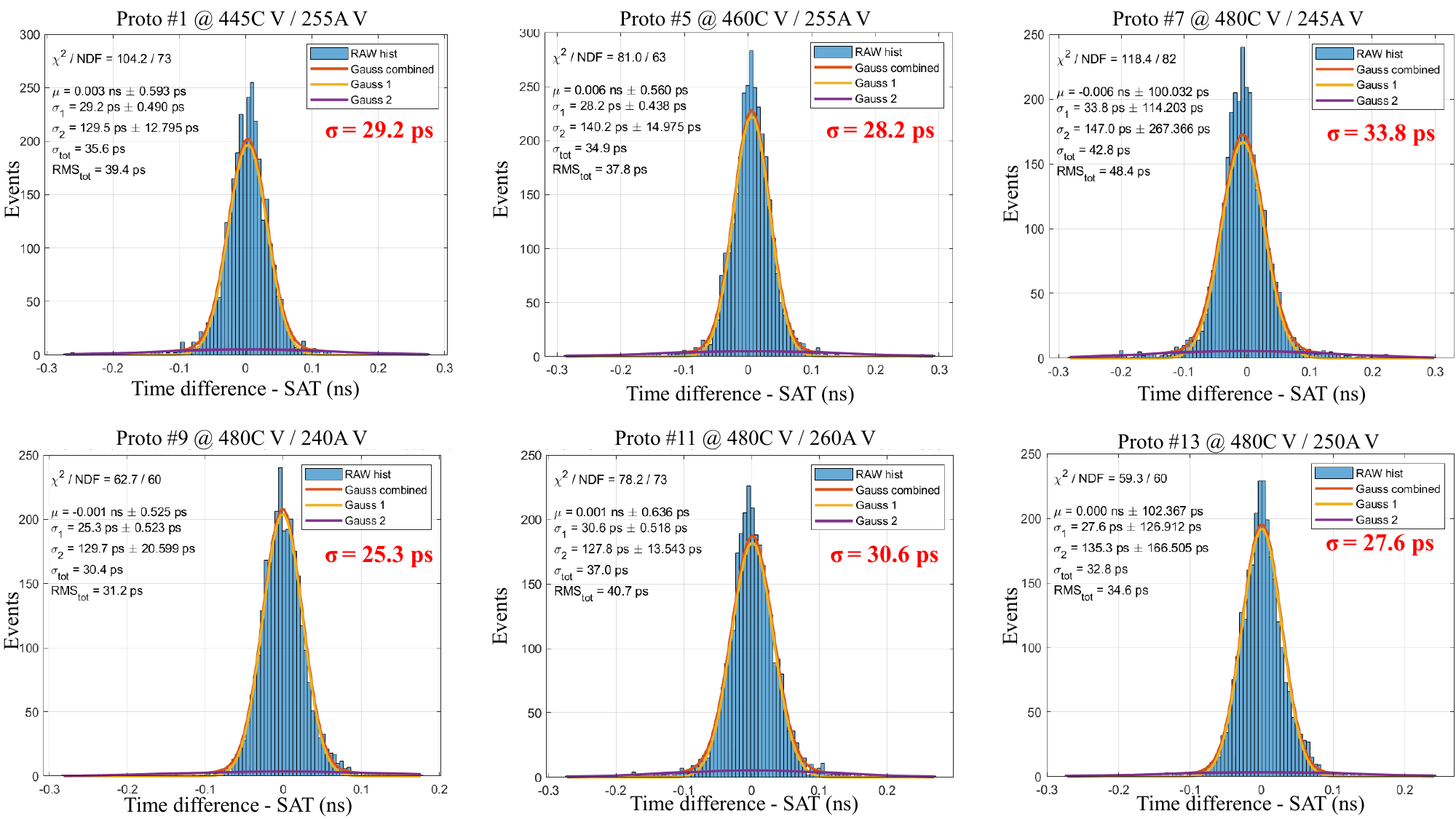}
\caption{\label{fig:timeResol} Distribution plots of the best time resolution measured for all six prototypes instrumented each with a CsI photocathode. The voltage settings for the anode and cathode of each prototype is reported in the title.}
\end{figure}

The best timing performance achieved for each of the six prototypes with a CsI photocathode during the July test beam is shown in the distribution plots of Fig.~\ref{fig:timeResol}. The voltages applied to the cathode and the anode are displayed in the title of each histogram. The test beam time period was too short for a full voltage scan study of all the prototypes. Therefore, the reported results represent the best time resolution for this test beam and not necessarily the optimal time resolution for each prototype. We believe that a comprehensive voltage scan study of the prototypes in the future will provide better timing performance. 
The best performance was achieved with proto~\#9 with a time resolution of 25.3~ps and the worst with proto~\#7 with a time resolution of 33.8~ps. Proto~\#7 and proto~\#11, both with hashed Cu readout pads, have a relatively poor time resolution ($\geq$30~ps). All other prototypes with a plain Cu readout pad have a time resolution better than 30~ps. This is an early indication that reducing the detector input capacitance does not seem to improve the timing performance. The best performance achieved with proto~\#5 at the July test beam was 28.2~ps with cathode and anode voltages of 460~V and 255~V respectively. This performance is significantly worse than the 23.5~ps measured with the same detector in the April test beam with cathode and anode voltages of 440~V and 255~V. The only difference between the two measurements is that different samples of the CsI photocathode were used. We suspect that photocathodes with lower quantum efficiency, due to the poor quality of the CsI layer deposition, were used during the July test beam resulting in a degraded time resolution. 
%
%
\subsection{Time resolution of various prototypes with a DLC photocathode}
\label{subsec:dlc_photoC} 
A reliable and repeatable high quality deposition of a CsI photocathode layer on the MgF$_2$ radiator substrates for PICOSEC applications is a difficult process, making the one-to-one comparison of the various prototypes with different parameters very challenging; each prototype used a different CsI photocathode. Moreover, because of the hygroscopic nature of CsI and its extreme sensitivity to moisture, the careful handling of this material during the assembly of the prototypes is essential, requiring minimum exposure to the environment even in a high quality clean room. In addition, CsI photocathodes are prone to rapid degradation when subjected to both ion backflow bombardment~\cite{mm-picosec-lisowska} and discharges which are unavoidable with gaseous detectors. These vulnerabilities make the use of CsI photocathodes extremely challenging for applications of $\muup$RWELL-PICOSEC technology in future HEP or NP experiments, especially for large size detectors.
\begin{figure}[!ht]
\centering
\includegraphics[width=0.995\columnwidth,trim={0pt 0mm 0pt 55mm},clip]{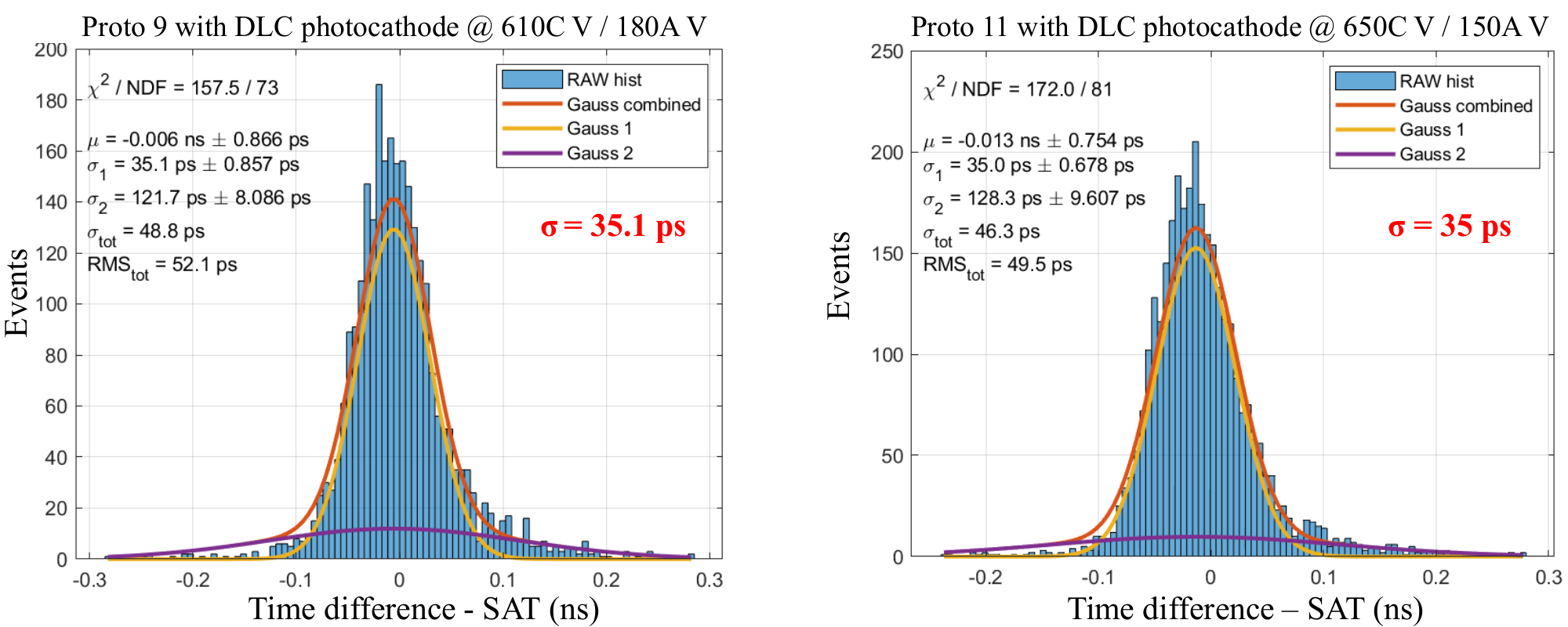}
\caption{\label{fig:DLC} Time difference distribution plots of proto~\#11 instrumented with a \textbf{(left)}: 18 nm CsI photocathode; \textbf{(right)}: 1.5 nm DLC photocathode.}
\end{figure}
For these reasons, a number of carbon-based alternative materials~\cite{mm-picosec-lisowska} with decent quantum efficiency, such as the Diamond-Like Carbon (DLC), Boron Carbide (B4C) and nano-diamond materials, are under investigation within the CERN PICOSEC collaboration to be used as more robust and easy-to-handle photocathodes PICOSEC detectors. A few single-pad $\muup$RWELL-PICOSEC prototypes with 1.5 nm thick DLC photocathodes were tested at the July test beam. The distribution plots of Fig.~\ref{fig:DLC} show the preliminary results of proto~\#9 (left) with a time resolution of 35.1~ps~$\pm$~0.86~ps at 610 V applied to the cathode and 180 V to the anode and of  proto~\#11 (right) with a time resolution of 31.~ps~$\pm$~0.68~ps at 650 V applied to the cathode and 150 V to the anode. The time resolution performance of the prototypes with DLC photocathodes is slightly worse than the one achieved with CsI photocathodes. This interesting observation confirms that the performance achieved with the CsI photocathodes is far from optimal and is due to the poor quality of the photocathodes. CsI photocathodes have higher quantum efficiency than DLC photocathodes resulting in twice as many photoelectrons, N$_{PE}$. This significantly improves the time resolution performance because N$_{PE}$ is the main parameter that contributes to the time resolution of PICOSEC detectors.  

%% file: 05_conclusion.tex
\section{Conclusion and Perspectives}
\label{sec:conclusion}
The $\muup$RWELL-PICOSEC detector concept, which is based on $\muup$RWELL~\cite{Bencivenni:2015} amplification device,  is the sister  technology of the MM-PICOSEC~\cite{mm-picosec-bortfeldt}, capable of delivering precision timing in the range of tens of picoseconds for TOF based particle identification applications in future HEP and NP experiments. The detector combines the $\muup$RWELL device for electron amplification in gas with  a Cerenkov radiator, coated on its inner side with a thin photocathode layer. The detector is coupled with fast and low noise pre-amplifier and high resolution digitizer for the signal readout. Several small single-pad $\muup$RWELL-PICOSEC prototypes were designed, fabricated as part of the Jefferson Lab LDRD program and successfully tested with the 150~GeV/$c$ muon beam at the CERN SPS North Area H4 beam line to demonstrate the proof of the concept of the technology. Studies of the optimization of time resolution performance and detector stability  were conducted on prototypes with various $\muup$RWELL hole geometrical parameters and different photocathode  materials. Initial results demonstrate that time resolution better than 24 ps is achieved. \\
Potential for  $\muup$RWELL-PICOSEC technology to achieve sub-20 ps time resolution with robust photocathodes as well as the scalability to large area (20~cm~$\times$~20~cm) detectors are under investigation. The next phase of R\&D effort on $\muup$RWELL-PICOSEC technology will focus on the development of precision timing and TOF applications for the muon detection system of the future circular collider (FCC-ee) at CERN.  

%% file: 06_acknowledgement.tex
\section{Acknowledgments}
\label{sect_acknowledgements}
The research described in this paper was conducted under the Laboratory Directed Research and Development (LDRD) Program at Thomas Jefferson National Accelerator Facility for the U.S. Department of Energy, Office of Science, and Office of Nuclear Physics under contract DE-AC05-06OR23177. We also acknowledge the  support of the CERN EP R\&D Strategic Program on Technologies for Future Experiments; the RD51 Collaboration, in the framework of RD51 common projects; the DRD1 Collaboration; the PHENIICS Doctoral School Program of Université Paris-Saclay, France; the Cross-Disciplinary Program on Instrumentation and Detection of CEA, the French Alternative Energies and Atomic Energy Commission; the French National Research Agency (ANR), project id ANR-21-CE31-0027; the Program of National Natural Science Foundation of China, grants number 11935014 and 12125505; the COFUND-FP-CERN-2014 program, grant number 665779; the Fundação para a~Ciência e~a Tecnologia (FCT), Portugal; the Enhanced Eurotalents program, PCOFUND-GA-2013-600382; the US CMS program under DOE contract No. DE-AC02-07CH11359;